\documentstyle[11pt]{article}
\begin{document}

{\Large Parametric S-tree Method and Its Generalizations}

{\bf Karen M.Bekarian} 

{\it Yerevan State University}

{\bf Anahit A.Melkonian}
  
{\it Yerevan Physics Institute}
\vspace{3mm}

{\bf Abstract.} 

A parametric approach  is developed to the method of S-tree diagrams
and its 
generalization  for
investigation of the hierarchical substructure of $N$-body nonlinearly
interacting systems (e.g., clusters of galaxies). 
The introduction of a parametric function allows us to 
take into account the individual features of the particles
which do not have a direct influence on the dynamics of the system
but are crucial for  analyzing  the output data. 
An algorithm is proposed for the latter based on
the construction of parametrical subgroups.

A generalization of the S-tree scheme  is presented as well.
This enables us to 
 perform parallel substructure analysis with respect to various
dependent or independent parameters, thus using fully the initial 
information of a system.

\section{Introduction}

Computers are powerful tools for the
investigation and numerical simulation of complex systems, including
information about
their hierarchical structures \cite{LM}. A class of such problems, namely
N-body gravitating configurations,
has an essential role in the understanding of various structures in the
universe such as star clusters, galaxies, clusters of galaxies, and so forth.

In the present paper we suggest an algorithm for
a statistical method of investigating the hierarchical structure
of gravitating N-body systems.  The large-scale distribution of 
galaxies in the universe and the hierarchical structure of clusters
of galaxies are two of the key
problems of  extragalactic astrophysics \cite{Col, Esc}.
 This is important because
the relaxation times of clusters of galaxies are of the order or
exceed the age of
the universe, those systems recall their initial conditions. 
Hence, the study of the hierarchical properties of systems
may directly indicate the
mechanisms of their formation and evolution.

Among the statistical methods used for  investigating  the
hierarchical
properties of  galaxy clusters are the
two-point correlation functions, Lee statistics, topological measures, 
wavelets and so on, for review see \cite{Peeb,GK}.

The studies already performed enable us to conclude that many galaxy clusters
may include subgroups with vivid dynamical peculiarities. The
existence of the subgroups is also apparent from
$X$-ray radiation data
 of the clusters. The development of more informative
methods for the analysis of the problem is of particular interest, 
given the increase of the possibilities of observational technique.

The $S$-tree diagram method \cite{GK, GHK} is based on the idea of 
introducing a {\it
degree of boundness} between the particles and their sets and therefore 
self-consistently 
using both  positional and kinematic information about the
particles.
The idea is therefore that the parameters of particles in a physically
interacting system, such as the components of their 3-velocities and
coordinates, have to
correlate with each other. On the contrary,  particles that do not
satisfy
the correlation have to be chance projections either in front of
or behind the physical system. The problem is therefore to reveal the
corresponding function of interaction, and to define its ``degree'', since
obviously, various particles can have different degrees of interaction.
This idea was realized {\it via} the S-tree technique, thus  revealing the
physically interacting system among the projected set of galaxies, and
moreover  establishing the subgroups with various degrees of interaction.
 This is achieved by  considering the geometrical properties of the
 phase space of the system, namely,  its two-dimensional curvature. The
N-body system as a hamiltonian system is thus being transferred {\it via} the 
Maupertuis principle to the
phase space and its dynamics are being described by the flow of geodesics
on a riemannian manifold with its metric determined by the potential of
interaction \cite{GS}. The S-tree method has already been  used for the study of
ESO Key Program data on the nearby Abell clusters \cite{GM2} and
many other observational data. 
In particular, the ESO data enabled the determination of a 
new class of galaxy configurations as dynamical entities given their
peculiar properties.

The S-tree technique has been developed not only for point
approximation, but also to account for the extension of  particles \cite{BM},
and to include certain generalized schemes \cite{BG}.

The parametric approach  of the S-tree method developed in the present paper allows us to
consider  individual properties of the particles, which are crucial
for the analysis of the output information on the hierarchical structure of the
system. 
For example, in the case of the galactic systems such a quantity can be the
morphological class,
which though does not influence the dynamics directly, has an
essential impact on  analyzing the output. The aim is therefore to have
the
algorithm include the parametric representation in the hierarchic
structure intrinsically.

Then we extend the approach to 
generalized schemes that enable the algorithm to use different functions 
of boundness self-consistently, thus
obtaining a more complete picture about the hierarchical structure
of the system.
We first introduce the necessary definitions and then move on to the algorithm.

\section{S-tree and its generalization}

We now briefly recall the main steps of the S-tree method.
As mentioned
previously, the key idea of this method is  the introduction of the
concept of the
degree of boundness $\rho$.

Consider a set of $N$ points:

$$
X= \{x_1, \dots, x_N \},
$$
the function $P$
$$
P: X \times X \rightarrow R_{+} \quad \rm{and} \quad \rho \in R_{+}.
$$
The basic definitions follow.

{\bf Definition 2.1.}
We say that $\forall x \in X$ and $\forall y \in X$ are $\rho$-{\it bounded}, if
$P(x,y) \geq \rho$.

{\bf Definition 2.2.}
We say  that $U \subset X (U \neq \emptyset)$ is a $\rho$-{\it bounded
subgroup}, if: 
\begin{enumerate}
\item  $\forall x \in U$ and $\forall y \in \bar U \Rightarrow P(x,y) < \rho$;

\item  $\forall x \in U$ and $\forall y \in U \quad \exists x=x_{i_1},
x_{i_2}, \dots,
x_{i_k}=y$, 

that $P(x_{i_l}, x_{i_{l+1}}) \geq \rho; \quad  \forall l=1, \dots,k-1$.
\end{enumerate}
{\bf Definition 2.3.}
We say  that $U_1, \dots, U_d$ is the {\it distribution} of the set $X$ {\it via}
$\rho$-bounded groups, if:
\begin{enumerate}
\item  $\bigcup_{i=1}^{d}U_i=X;$

\item $i \neq j \quad (i,j=1, \dots, d) \Rightarrow U_i \bigcap U_j= \emptyset;$

\item $U_d (i=1,\dots,d)$ is a $\rho$-bounded group.
\end{enumerate}
It is possible for the function $P$  to consider different physical
quantities.
In \cite{GK}, various quantities such as the energy, the potential,
perturbations of
potential, momentum, and so on,  are listed and discussed. For example, the
choice
of the mutual distance of the particles to define the subgrouping,
obviously,
is equivalent to the corresponding correlation functions
(spatial or angular, related with each other simply {\it via} the Limber equation),
which is a rather incomplete characteristic for that aim.
 It can be shown that, at least for astrophysical problems, among the most
informative ones is the riemannian curvature of the configuration space
\cite{GS},
which determines the behavior of close geodesics, as known from  basic
courses on classical mechanics \cite{Arn}.

So, by the S-tree algorithm we obtained the distribution of the N-body system for
any given function $P$ and $\forall \rho$. This distribution will satisfy
Definitions 2.1, 2.2 and  2.3. The final result can be
represented either through
tables or by graphs (S-tree).     
 
Now we aim to use the S-tree scheme for different
boundness criteria simultaneously. 

We have $t$ different functions
$P_1, \dots, P_t$, which satisfy the definitions  introduced in the method
of S-diagrams  and that describe the degree of
interaction of the system's
members. To each $P_{\alpha}$ we correspond a matrix $D_{\alpha}$, where
$D_{\alpha}=(d_{ij}^{\alpha}), \alpha=1, \dots, t$ \cite{BG}.

Consider the following matrix $D$:

$$
D=(\bar d_{ij}), \quad i,j=1, \dots, N,
$$
where 
$$
\bar d_{ij}=(d_{ij}^1, d_{ij}^2, \dots, d_{ij}^t), \quad i,j=1, \dots, N,
$$
which  contains the entire information of the system, that is, it  uses all 
$P_t$ functions. For example, the previously mentioned Riemannian curvature is
a function of the coordinates, velocities of all the particles, their
masses, potential of interaction, its derivatives, and so on.     

Each function $P_{\alpha}, \alpha=1,\dots, t$ corresponds to a definite
vector of boundness $\bar \rho_{\alpha}=(\rho_1^{\alpha}, \dots, \rho_{
 Q_{\alpha}}^{\alpha})$  and $Q_{\alpha}$
satisfies the conditions of the choice of the vector of boundness
\cite{GK}.

We construct the vector $\bar \rho_D=(\bar \rho_1^D, \dots, \bar
\rho_{Q_D}^D),$ where
$\bar \rho_k^D=(\rho_{k_1}^1, \dots, \rho_{k_t}^t);$ 

$\quad k=1, \dots, Q_D; \quad k_{\alpha}=1, \dots, Q_{\alpha}, \quad
\alpha=1, \dots, t$.

The vector $\bar \rho_D$ is obtained with the help of all possible variations
of the components of $\bar \rho_k^D (\alpha=1, \dots, t; k_{\alpha}=1,
\dots, Q_{\alpha}),$ and it is obvious  that $Q_D \leq ((N^2-N)/2+1)^t$.

Taking into account the fact that different $P_{\alpha}$ functions contain
different information about the system, we introduce the concept of
the ``degree of influence'' $E_{\alpha}$ for each of the functions under
consideration. We will conditionally understand  $E_{\alpha}$ as the number of
physical quantities 
(distance, potential of the  interaction 
and its derivatives, and so on), existing in the $P_{\alpha}$ formulae.

The next step of the algorithm is the transition from matrix $D$  to matrix $D_u$
with the help of a fixed current vector $\bar \rho_k^D$.

$$
D_u=(\bar u_{ij}); \quad i,j=1, \dots, N,
$$
where $\bar u_{ij}=(u_{ij}^1, \dots, u_{ij}^t); \quad u_{ij}^{\alpha}
(\alpha=1, \dots, t)$ is defined in the following way:

$$
u_{ij}^{\alpha}=\left\{ \begin{array}{rl}
             E_{\alpha}&{\rm if}\quad d_{ij}^ \alpha \geq
                       \rho_{k_{\alpha}}^{\alpha} \\
             0             &{\rm if}\quad d_{ij}^ \alpha <
                           \rho_{k_{\alpha}}^{\alpha}.
                \end{array}
                \right.
$$

Thus, the matrix $D_u$ contains  information on the degree of influence
of each criterion too. Then we construct the matrix $D_v$  in
the following way:

$$
D_v=(v_{ij}); \quad i,j=1, \dots, N,
$$
where $v_{ij}=\sum_{\alpha=1}^{t}u_{ij}^{\alpha}.$

The matrix $D_v$  characterizes the quantitative degree of boundness. At
the same time we also take into account   information on the
$\rho$-boundness of particles, because matrix $D_v$  is defined with the
help
of matrix $D_u$. According to the S-tree method we then construct the
vector of boundness $\bar \mu$ for the matrix $D_v$:

$$
\bar \mu=(\mu_1, \dots, \mu_T).
$$
It is obvious  that $T \leq ((N^2-N)/2+1)$. 

Note, that while estimating
$Q_D$ in this particular case we are taking into account the symmetry of
the initial matrices. 

Thus  the N-body system, according to the generalized S-tree method, is split
into $\tilde
\rho$-bounded subgroups, where
$\tilde \rho=(\bar \rho_k^D; \mu_l);  (l=1, \dots, T)$ in the
following way.  We obtained matrix $D_u$  from matrix  $D$  with the help of some
fixed $\bar \rho_k^D$, afterwards we get matrix $D_v$ out of $D_u$ using the described method.
The last step of the algorithm  is the distribution of the system into subgroups 
{it via} matrix $D_v$  for the fixed
$\mu_l$ using the S-tree method.

To get complete information of the distribution it is necessary to apply the
algorithm for all the possible pairs $(\bar \rho_k^D; \mu_l)$, where $\bar
\rho_k^D$ and $\mu_l$ accordingly are the components of the vectors $\bar 
\rho_D$ and $\bar \mu$.

The time $L$, which is needed for the realization of this algorithm, is
$L \leq ((N^2-N)/2+1)^{t+1}$ and for $t<<N$ no difficulties can arise
\cite{Cof}.

\section{Parametrical approach} 

As  mentioned previously, the S-tree method is 
splitting the system  into subgroups 
according to  the boundness parameter $\rho$. 

The application of a parametrical approach to the S-tree method has two main goals
\begin{enumerate}

\item To have a convenient way of representing output information of
the
S-tree algorithm with respect to individual properties
of the particles.

\item To have  more complete and precise information about the
substructure in the output information.
\end{enumerate}
Two possible versions of parametrical approach are considered.

1. After obtaining the distribution of the system by subgroups (by the
S-tree method), we introduce a parameter describing
certain  properties of particles independent of the
distribution.

Consider the set of $N$ discrete points

$$
X=\{x_1, \dots, x_N \}
$$   
and the function $P$, where

$$
P: X \times X \rightarrow R_+.
$$

Let us introduce the following function

$ B: X \rightarrow X \times R;$

$B(x_i)=(x_i, b_i).$
We denote $(x_i,b_i)=x_i^{b_i}$ for convenience, where $b_i$ is a 
parameter of $P$ depending on the system under consideration.

2. The suggested method is applicable if these  constraints are
satisfied simultaneously:
\begin{enumerate}
\item  $k \neq N,$ where $k$ is the number of subgroups;

\item  $k \neq 1$ and $\exists l_i, l_j,$ such that
$t_{l_i} >1$ and $t_{l_j} >1,$ where $t_{l_c}$ is the number of particles in the 
$l_c$-th subgroup, $l_c=1, \dots, k$.
\end{enumerate}
For any fixed $x_i$ the following sum is considered:
 $$P_i^d= \Sigma P(x_i,x_q),$$
where  $d$ is the number of $U_{j, \dots, k}$ subgroups and the
summation is performed for all $q=j, \dots, k$ (accordingly, $x_j, \dots, 
x_k$ forms the subgroup $U_{j, \dots, k}$ and $x_i \not \in U_{j,
\dots, k}$).

The sum 
$$P_i^0= \Sigma P(x_i, x_q)$$
(the summation is performed for all
$q=j, \dots, k$, where $x_i, x_j, \dots, x_k$ form the subgroup $U_{i,
\dots, k}$)  is considered as well.
 The concept of parametric boundness is introduced in the Definition 3.1.

{\bf Definition 3.1.} 
If $P_i^0$ is more than any  other $P_i^d$, than $x_i$ is called a
``parametrically real" (p.r) member of the subgroup; otherwise it is called
 ``parametrically imaginary" (p.i).

The final result of the output information in the parametric
approximation is presented in the following way.
If $x_i$ is a p.r. member of subgroup $U_{l_i}$, we denote $x_i$
as $x_i^{l_i}$;
   if $x_i$ is a p.i. member of subgroup $U_{l_i}$, we denote $x_i$
as $x_i^{l_j}$, where $l_j$ is the number of the subgroup $U_{l_j}$ with
elements $x_j, \dots, x_k$ on which the function $P_i^d (d=l_j)$  obtains its
maximal value.

\section{Generalized parametric scheme}

Now we consider the corresponding generalizations for  both
versions of the parametrical approach.

I. Consider the set of N points:
$X= \{x_1, \dots, x_N \}$ and the function $P$, where
$$
P: X \times X \rightarrow R_+.
$$

For the obtained distribution ({\it via} the  S-tree method) we introduce a
multidimensional parameter, each component of which shows a definite
peculiarity of the element, independent of the distribution.

Consider the following function
$$\begin{array}{rl}
B: X \rightarrow X \times R^k, k=&1,2, \dots\\
&\\
B(x_i)=&(x_i,\bar b_i),
\end{array}$$
where $\bar b_i=(b_i^1, \dots, b_i^k).$

We introduce the  notation $(x_i,\bar
b_i)=x_i^{\bar b_i}$ for convenience, where $b_i^j; j=1, \dots, k$ is not
a variable of $P$.

In this particular case $\bar b_i$ is a multidimensional parameter, each
component of which is normalized in accord with the physical content of
the corresponding component and the system under investigation.

II. The initial limitations of the proposed method fully correspond to the
limitations of the parametrical approach.

We fix some value $x_i$ and  consider all possible sums $P_i^d=
\Sigma P(x_i, x_q),$ where $d$ is the number of the subgroup $U_{j,
\dots, k}$ and the summation is taken for all the $q=j, \dots, k$
(accordingly $x_j, \dots, x_k$ forms the group $U_{j, \dots, k}$ and $x_i
\not \in U_{j, \dots, k})$.

It is obvious  that the vector $\bar P_i^d=(P_i^{d_1}, \dots,
P_i^{d_{k-z}})$, where $k-z$ is the number of those subgroups for which
the above mentioned limitations have been carried out.

We construct the vector $\tilde P_i^d$  from vector $\bar P_i^d$, where 
$\tilde P_i^d=(P_i^{d_{j_1}}, \dots, P_i^{d_{j_{k-z}}})$, the components
of which are the components of the vector $\bar P_i^d$ situated in 
decreasing order.

Let us consider the sum 
$$P_i^0= \Sigma P(x_i,x_q)$$,
(the summation is 
taken by all $q=j, \dots, k$), where $x_i, x_j, \dots, x_k$ forms the
group $U_{i, \dots, k}$. 

The concept of the generalized parametric
boundness is introduced in Definition 4.1.

{\bf Definition 4.1}
If $P_i^0$ is greater then all components of the vector $\bar P_i^d$, 
we  say
that $x_i$ is ``parametrically real" (p.r.) member of the group $U_{i,
\dots, k}$, otherwise $x_i$  is ``parametrically imaginary" (p.i.) member of
the group $U_{i, \dots, k}$.

The final representation of the output information in terms of generalized parametric approach
is made in the following way:

if $x_i$ is a p.r. member of the group $U_{l_i}$,  denote $x_i$ as
$x_i^{l_i}$;

if $x_i$ is a p.i. member of the group $U_{l_i}$,  denote $x_i$ as
$x_i^{\bar l_j}$, where $\bar l_j=(l_{j_1}, l_{j_2}, \dots)$. $l_{j_r}$
are numbers of those subgroups, where $P_i^{d_{j_r}} > P_i^0;$ 
$r=1, \dots, k-z; \quad d_{j_r}=l_{j_r}$.

\section{Conclusion}

In this paper a parametric approach to the S-tree diagram method 
has been
suggested. The proposed approach is useful for revealing the optimal
distribution of the system under investigation \cite{BM}.

The following developments were presented.
\begin{itemize}
\item A generalization of the discrete S-tree method.

\item A generalization of the parametric approach to the S-tree method , in the case of
multiparametric functions.
\end{itemize}
The methods described  are more informative since they use all
the  information of the system in the input, while  they 
enable the output to have a hierarchical structure depending not only on the 
kinematic and positional data, but also on the parameters having
no direct influence on the dynamics of the system.

For example, in the case of clusters of galaxies the algorithm will use
data from the catalogues not only on the coordinates and redshifts of
galaxies but also the morphological, color, spectral index, and other
information 
self-consistently in the output information on the hierarchic
structure of the systems.

The hierarchical structure of N-body nonlinearly interacting particles
is only one indication of their complex dynamics \cite {Co},\cite{GMO}.

{\bf Acknowledgments}

M.A. is grateful to Prof. S. Randjbar-Daemi for
hospitality in ICTP.
We  thank  Prof. V. Gurzadyan for valuable discussions 
and  A. Sanasarian for assistance.

\end{document}